\documentclass[reprint,twocolumn,nofootinbib,floatfix]{revtex4-1}
\usepackage{graphicx}
\usepackage[utf8]{inputenc}
\usepackage[T1]{fontenc}

\usepackage{braket}
\usepackage{amsmath}
\usepackage{amssymb}
\usepackage{csquotes}
\usepackage{bm}
\usepackage{graphicx}
\usepackage{siunitx}
\usepackage{notes2bib}
\usepackage[svgnames,x11names]{xcolor}

\bibnotesetup{note-name = ,
use-sort-key = false}

\def\BZ{\ensuremath{\text{BZ}}}

\def\Id{\ensuremath{\text{Id}}}
\def\dd{\ensuremath{\mathrm{d}}}
\def\ee{\ensuremath{\mathrm{e}}}
\def\ii{\ensuremath{\mathrm{i}}}

\def\tr{\ensuremath{\operatorname{tr}}}
\def\deg{\ensuremath{\operatorname{deg}}}

\usepackage[breaklinks=true,colorlinks=true,linkcolor=blue,urlcolor=blue,citecolor=blue]{hyperref}
\usepackage{bookmark}

\begin{document}
\title{Complex classes of periodically driven topological lattice systems}
\author{Michel Fruchart}
\email{michel.fruchart@ens-lyon.fr}
\affiliation{Laboratoire de Physique, École Normale Supérieure de Lyon, 47 allée d'Italie, 69007 Lyon, France}
\date{21 March 2016}

\begin{abstract}
Periodically driven (Floquet) crystals are described by their quasi-energy spectrum. Their topological properties are characterized by invariants attached to the gaps of this spectrum. In this article, we define such invariants in all space dimensions, both in the case where no symmetry is present and in the case where the unitary chiral symmetry is present. When no symmetry is present, a $\mathbb{Z}$-valued invariant can be defined in each gap in all even space dimensions. This invariant does not capture all the properties of a system where chiral symmetry is present. In even space dimension, chiral symmetry puts constraints on its values in different gaps. In odd space dimension, chiral symmetry also enables to define a $\mathbb{Z}$-valued invariant, only in the chiral gaps $0$ and $\pi$. We relate both gap invariants to the standard invariants characterizing the quasi-energy bands of the system. Examples in one and three dimensions are discussed.
\end{abstract}


\maketitle

\section*{Introduction}

Topological properties of waves in a spatially periodic medium (e.g. electrons in a crystal) are characterized by so-called topological invariants, the nature of which depends on the symmetries of the system. Soon after the discovery of time-reversal invariant topological insulators by Kane and Mele in 2005 \cite{KaneMele2005}, a milestone in the understanding of topological phases was set by the discovery of a relationship between the Altland-Zirnbauer classes (and the corresponding non-spatial discrete symmetries) and the possible topological invariants of a given system, leading to the so-called periodic table of topological insulators \cite{Kitaev2009,SchnyderRyuFurusakiLudwig2009,RyuSchnyderFurusakiLudwig2010}.
In parallel, it was proposed by several authors to use a time-periodic excitation to induce topological transitions, with the goal of creating tunable materials, the so-called Floquet topological insulators \cite{OkaAoki2009,InoueTanaka2010,LindnerRefaelGalitski2011,KitagawaOkaBrataasFuDemler2011}. Despite being out-of-equilibrium, Floquet topological phases share a lot of similarities with equilibrium ones, and as long as phase coherence is preserved by the driving field as well as by the inevitable dissipative processes, they should be well described by an effective unitary evolution.
Yet, Floquet topological phases also display peculiar properties, as was first discovered by Kitagawa and collaborators \cite{KitagawaBergRudnerDemler2010,KitagawaBroomeFedrizziRudnerBergKassalAspuru-GuzikDemlerWhite2012}, both theoretically and experimentally. To fully describe the topological properties of the effective unitary evolution of a two-dimensional periodically driven system
without specific symmetry, Rudner, Lindner, Berg and Levin \cite{RudnerLindnerBergLevin2013} developed a new framework, which essentially consists in computing the winding number of the periodic part of the evolution operator of the system. When symmetries are present, constraints are set on the periodized evolution operator and change the nature of the possible topological properties, in the same way that the topological properties of an energy band (represented by a spectral projector) depend on the symmetry constraints it must satisfy. For example, when time-reversal symmetry is present, a $\mathbb{Z}_2$-valued index characterizes each gap \cite{CarpentierDelplaceFruchartGawedzki2015}. Several steps towards a full classification of periodically driven topological phases were recently taken by Nathan and Rudner \cite{NathanRudner2015}, who identified the origin of anomalous behavior of Floquet systems (where equilibrium invariants are not sufficient) as stemming from topologically protected crossings in the time-dependent spectrum of the evolution operator, and by Fulga and Maksymenko \cite{FulgaMaksymenko2015} who started to extend to periodically driven systems the scattering approach already used to characterize static system \cite{FulgaHasslerAkhmerovBeenakker2011,FulgaHasslerAkhmerov2012}.

Here, we aim at defining bulk topological invariants characterizing periodically driven systems without any symmetry and with chiral symmetry, in any dimension. This corresponds to classes A and AIII of the periodic table of topological phases, which are called complex classes as they involve complex K-theory. Both classes were previously studied respectively in two dimensions by Rudner and collaborators \cite{RudnerLindnerBergLevin2013} and in one dimension by Asbóth and collaborators \cite{AsbothTarasinskiDelplace2014}. The case without symmetry is straightforwardly extended to any even dimension, and only the relationship between this invariant and the Chern invariant needs attention. When chiral symmetry is present, we take a different route to defining the topological invariants by starting directly from the periodized evolution operators: although the results are the same, this approach to defining gap invariants is more easy to extend to any (odd) dimension and to relate to static invariants. In particular, a new gap invariant for chiral symmetric systems in three dimensions is introduced, and illustrated on a simple driven tight-binding model. In addition to being interesting from the fundamental point of view, topological invariants in higher dimensions are relevant in quasi-crystals \cite{Kraus2012,Kraus2013,Prodan2015} and in systems with artificial/synthetic dimensions \cite{OzawaPriceGoldmanZilberbergCarusotto2015}.

\section{An example with chiral symmetry in 1D: the driven SSH model}
\label{example_chiral}

\begin{figure}[t]
	\centering
	\includegraphics{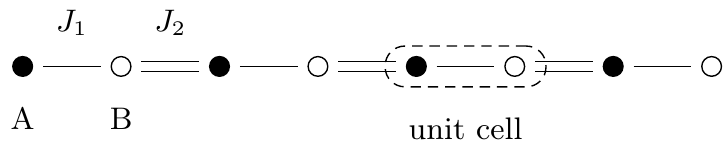}
	\vspace*{-0.25cm}
	\caption{\label{ssh_lattice} Lattice of the (driven) SSH model.}
\end{figure}

Consider a simple well-known example \cite{GomezLeonPlatero2013,AsbothTarasinskiDelplace2014,DalLagoAtalaFoaTorres2015}, the driven SSH (Su-Schrieffer-Heeger, see \cite{SuSchriefferHeeger1979}) model, a one-dimensional system with two sublattices $A$ and $B$ (see figure~\ref{ssh_lattice}) described by the tight-binding Hamiltonian
\begin{equation}
\begin{split}
	H(t) = \sum_{x} &J_1(t) \ket{x,A}\bra{x,B} + \\
					&J_2(t) \ket{x,A}\bra{x+1,B} + \text{h.c.}
\end{split}
\end{equation}
where the sum runs on points $x$ of the Bravais lattice, $A$ and $B$ represent the two sublattices of the crystal, and $\ket{x,A/B}$ represents a state localized at point $x$ on the sublattice $A/B$. This corresponds to the Bloch Hamiltonian 
\begin{equation}
	H(t,k) = (J_1(t) + J_2(t) \cos(k)) \sigma_1  + J_2(t) \sin(k) \sigma_2
\end{equation}
where $\sigma_i$ are the Pauli matrices in the basis of sublattices $A$ and $B$. We choose $J_1(t) = J_1 + A \cos(\omega t)$ and $J_2(t) = J_2$. Chiral symmetry is represented by the operator $\Gamma = \sigma_3$, which anticommutes with the Hamiltonian whilst reversing time, $\Gamma H(t,k) \Gamma = - H(-t,k)$.

The undriven system with $A=0$ is a band insulator when $|J_2/J_1| \neq 1$, and it is topologically nontrivial when $|J_2/J_1| > 1$. At the edge of a finite nontrivial system appear topologically protected zero modes (with an energy inside the bulk gap) exponentially located close to the boundary. In contrast, the edge of a trivial system does not host zero modes. 

\begin{figure}[!t]
	\centering
	\includegraphics{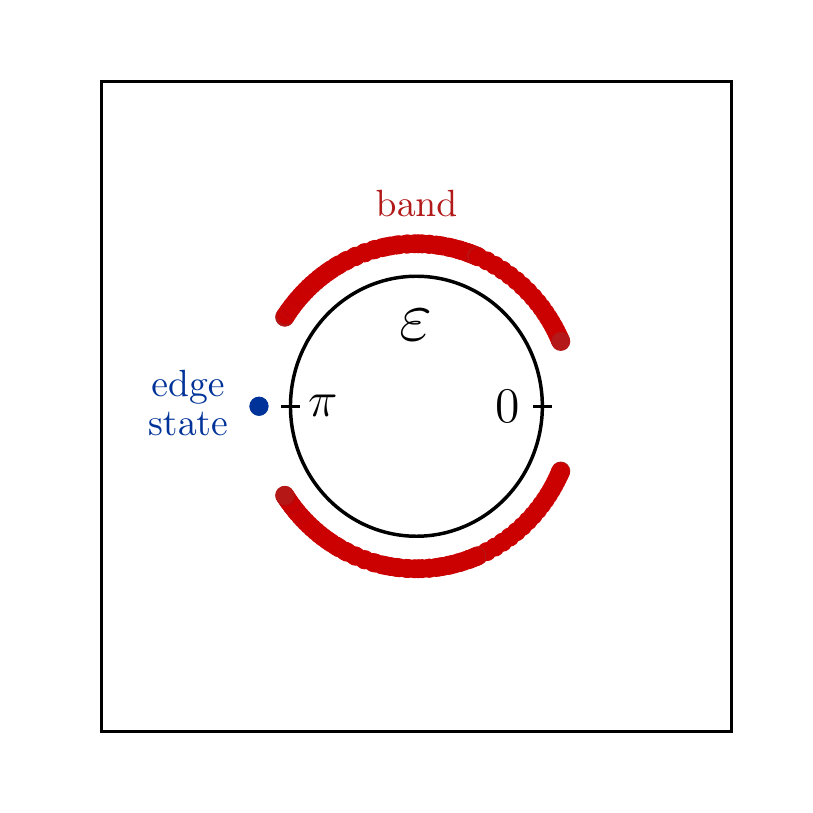}
	\vspace*{-0.5cm}
	\caption{\label{figure_ssh_dynamic_pi_nontrivial} Quasi-energy spectrum of a finite system (with edges) of the driven SSH model for $J_1 = \num{3/2}$, $J_2 = \num{1}$, $A = \num{6}$ and $T = \num{2.35}$. The undriven system with $A = \num{0}$ is trivial, but the driven phase at this point of the parameter space the driven system is topologically nontrivial, as $G_{\pi}[U] = \num{1}$ (and $G_{0}[U] = \num{0}$ as this gap is trivial). As a consequence, edge states appear in the quasi-energy spectrum of the finite system (in blue, as opposed to the bulk bands which are in red), only in gap $\pi$, and can be related to the static chiral invariant of the effective Hamiltonian. The quasi-energy spectrum is obtained by diagonalization in the Sambe space truncated to \num{19} sidebands for a system of length \num{80}. The invariants were computed from the bulk Hamiltonian by direct integration. The code used for all computations is available in Ref.~\bibnotemark[sourcecode] .
	}	
\end{figure}

Turning on the driving from a trivial point of the equilibrium phase diagram, say $J_1 = \num{3/2}$ and $J_2 = \num{1}$, one can bring the system to an out-of equilibrium phase where edge states appear inside the bulk gap, as in the equilibrium case, as illustrated in figure~\ref{figure_ssh_dynamic_pi_nontrivial}. These properties can be explained by computing the standard chiral invariant from the effective Hamiltonian $H^{\text{eff}}(k) = \ii/T \log U(T, k)$. A notable difference with equilibrium systems is that in periodically driven systems, the energy spectrum is no longer relevant, as energy is (explicitly) exchanged with the driving field and (implicitly) with the environment. However, because of the periodicity of the driving, energy modulo $\hbar \omega$ (where $\omega = 2 \pi/T$ is the driving angular frequency), a quantity called quasi-energy, is still conserved. From this definition, quasi-energy is a periodic quantity, that is, $\epsilon$ and $\epsilon + n \hbar \omega$ with an integer $n$ represent the same physical quantity. The band spectrum therefore organizes on a circle, and a system with two quasi-energy bands has not one but two gaps, both of which may host topological edge states. 

\begin{figure}[!t]
	\centering
	\includegraphics{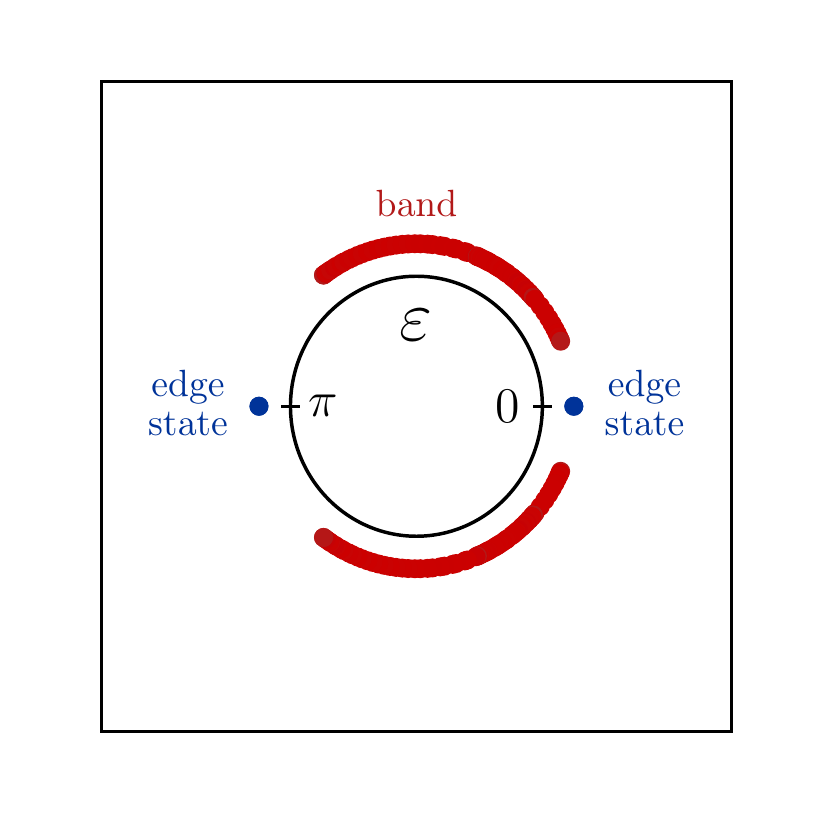}
	\vspace*{-0.5cm}
	\caption{\label{figure_ssh_dynamic_anomalous} Quasi-energy spectrum of a finite system (with edges) of the driven SSH model for $J_1 = \num{1}$, $J_2 = \num{3/2}$, $A = \num{6}$ and $T = \num{2.35}$. In this point of the parameter space, the driven system is topological and anomalous, as $G_{0}[U] = G_{\pi}[U] = \num{1}$. As a consequence, edge states appear in the quasi-energy spectrum of the finite system (in blue, as opposed to the bulk bands which are in red), both in gap $0$ and in gap $\pi$. See figure~\ref{figure_ssh_dynamic_pi_nontrivial} for details on the computation.
	}
\end{figure}

A striking property of Floquet systems is the existence of so-called anomalous topological systems \cite{KitagawaBergRudnerDemler2010,JiangKitagawaAliceaAkhmerovPekkerRefaelCiracDemlerLukinZoller2011,RudnerLindnerBergLevin2013}, where all the topological invariants characterizing the effective Hamiltonian vanish, and yet the out-of-equilibrium phase hosts topologically protected states at its boundary, related to a more subtle bulk invariant. This situation is illustrated in figure~\ref{figure_ssh_dynamic_anomalous}. Here, both gaps host topological edge states, which illustrates the necessity of a bulk invariant more precise than the standard chiral invariant applied to the effective Hamiltonian, and that characterizes a quasi-energy gap, and not a band. As we will see, only the gaps corresponding to quasi-energies $1 = \ee^{\ii 0}$ and $-1 = \ee^{\ii \pi}$ can host edge states topologically protected by chiral symmetry. Alternatively, chiral bulk invariants may only be defined for two quasi-energies $\varepsilon = 0$ and $\pi$, so only two indexes $G_{0}$ and $G_{\pi}$, both $\mathbb{Z}$-valued, will be defined. In the current example, both are equal and nonzero : $G_{0} = \num{1} = G_{\pi}$.

\bibnotetext[sourcecode]{See Supplemental Material at \url{http://arxiv.org/src/1511.06755/anc} for the code used to produce the figures and in the numerical computation of the topological invariants in the examples.}

\section{Floquet theory}

Floquet theory enables to separate slow and fast parts of the dynamics (with respect to the driving period), which are respectively encoded in the Floquet operator $U(T)$ and in the evolution operator over one period, $[0,T] \ni t \mapsto U(t)$, or more precisely in its periodized version $V(t)$ (to be defined), which can be Fourier transformed.

Consider a (one particle) Hamiltonian periodic in time, 
\begin{equation}
	H(t+T) = H(t).
\end{equation}
Define the evolution operator $U(t,t')$ as the solution to the Cauchy problem $\ii \partial_t U(t,t') = H(t) U(t,t')$ with initial value $U(t',t') = \Id$. The periodicity of the Hamiltonian is equivalent, in terms of the evolution operator, to
\begin{equation}
	U(t+T,t'+T) = U(t,t').
\end{equation}
In the following, we will choose an origin of time and set $U(t)=U(t,0)$ to simplify the notations. When the Hamiltonian is periodic in time, Floquet theory ensures that the evolution operator is \enquote{quasi-periodic} in time, that is
\begin{equation}
	U(t + n T) = U(t) \left[ U(T) \right]^n.
\end{equation}
The evolution operator after one period, $U(T)$, is called the Floquet operator, and its eigenvalues form the quasi-energy spectrum of the system (represented in figures \ref{figure_ssh_dynamic_pi_nontrivial} and \ref{figure_ssh_dynamic_anomalous} for the example developed in section \ref{example_chiral}).

We are interested in spatially periodic media, in which Bloch theorem allows a decomposition of the single-particle Hamiltonian into a family of so-called Bloch Hamiltonians $H(k)$, where $k$ is the quasi-momentum living on the Brillouin torus~$\BZ$, which has the same dimension as the space. It is always possible to choose $H(k)$ so it is periodic in $k$, or equivalently so $H$ is a smooth function over the torus. In this scheme, the evolution operator also depends on $k$, and is computed pointwise from $H(k)$.

Suppose that there is a gap in the spectrum of the Floquet operator $U(T)$ around the value $\ee^{- \ii T \epsilon}$ (which will be designated as $\epsilon$ to shorten), that is, for all $k \in \BZ$, $\ee^{- \ii T \epsilon}$ is not in the spectrum of $U(T,k)$. One can then define an effective Hamiltonian at~$\varepsilon = T \epsilon$ (here $\varepsilon$ is a dimensionless quasi-energy which lies e.g. from $0$ to $2\pi$ or from $-\pi$ to $\pi$)
\begin{equation}
	\label{Heff_definition}
	H^{\text{eff}}_{\varepsilon}(k) = \frac{\ii}{T} \log_{-\varepsilon} U(T, k)
\end{equation}
where $\log_\alpha$ is the complex logarithm with cut along an axis with angle~$\alpha$, defined as
\begin{equation}
	\label{def_log}
	\log_{-\varepsilon}(\ee^{\ii \varphi}) = \ii \varphi
	\qquad
	\text{for}
	\qquad
	- \varepsilon - 2 \pi < \varphi < - \varepsilon.
\end{equation}
More precisely, the effective Hamiltonian is defined through the spectral decomposition 
\begin{equation}
	U(T,k) = \sum_{j} \lambda_j(k) \ket{\psi_j(k)}\,\bra{\psi_j(k)}
\end{equation}
by setting
\begin{equation}
	H^{\text{eff}}_{\varepsilon}(k) = \frac{\ii}{T} \sum_{j} \log_{- \varepsilon}(\lambda_j(k)) \ket{\psi_j(k)}\,\bra{\psi_j(k)}
\end{equation}
With a slight abuse of language, the spectrum of the effective Hamiltonian may also be called the quasi-energy spectrum of the system, as it is naturally related to the previously defined quasi-energy spectrum by the exponential.
Now define the periodized evolution operator
\begin{equation}
	\label{periodized_evolution_operator_def}
	V_{\varepsilon}(t,k) = U(t,k) \ee^{\ii t H^{\text{eff}}_{\varepsilon}(k)}
\end{equation}
so that $V_{\varepsilon}(t+T,k) = V_{\varepsilon}(t,k)$ and in particular $V_{\varepsilon}(0,k) = \Id = V_{\varepsilon}(T,k)$ for all $k$.

\section{Class A : without symmetries}
\label{sec:classA}

The simplest case happens when all symmetries are broken, that is, there are no constraint on the Hamiltonian.

When the dimension $d$ of the space (and therefore of the Brillouin torus $\BZ$) is even, $d=2n$, then $V_{\varepsilon}$ is a map from the odd-dimensional manifold $S^1 \times \BZ$ to the unitary group $U(N)$ (where $S^1$ represents a period of time, e.g. $[0,T]$), so we can define a $\mathbb{Z}$-valued invariant, extending the definition of \cite{RudnerLindnerBergLevin2013} to all even space dimensions by
\begin{equation}
	W_{\varepsilon}[U] = \deg(V_{\varepsilon}) \in \mathbb{Z}.
\end{equation}
where the degree (or winding) of a map $V$ from a $(2n+1)$-dimensional closed manifold $M$ to the unitary group $U(N)$ is defined \cite{BottSeeley1978} as\footnote{We usually expect the winding numbers \eqref{def_winding} of an unitary map to be an integer. This is only guaranteed when $M$ is a sphere \cite{BottSeeley1978}. In our case, $M$ is a $d+1$-dimensional torus, so whenever all lower-dimensional windings of a map $U : M \to U(N)$ vanish, $U$ can be deformed to another map, which can be seen as actually defined on the sphere (by contracting lower-dimensional cycles). In this case (when weak invariants vanish), the top-dimensional winding (strong invariant) is an integer. In general, it may not be an integer, but this does not prevent it from being a proper topological invariant.}
\begin{equation}
	\label{def_winding}
	\deg(V) = \deg_{2n+1}(V) = \int_{M} \omega_{2n+1}[V] \in \mathbb{Z}
\end{equation}
where
\begin{equation}
	\omega_{2n+1}[V] = (-1)^n \frac{n!}{(2n+1)!} \, \left(\frac{\ii}{2 \pi}\right)^{n+1} \tr\left[ (V^{-1} \dd V)^{2n+1} \right].
\end{equation}
It counts the number of times the map $V$ winds around nontrivial $(2n+1)$-cycles in $U(N)$.
The power should be understood as the exterior power of differential forms, so in terms of the derivatives of $V$, we integrate
\begin{equation}
\begin{split}
	&\omega_{2n+1}[V] = (-1)^n \frac{n!}{(2n+1)!} \, \left(\frac{\ii}{2 \pi}\right)^{n+1} \epsilon^{i_1 i_2 \cdots i_{2n+1}}
	\\
	& \tr\left[  (V^{-1} \partial_{i_1} V) \cdots (V^{-1} \partial_{i_{2n+1}} V) \right] \, \dd x_1 \dots \dd x_{2n+1}.
\end{split}
\end{equation}
where $\epsilon$ is the Levi-Civita symbol. For instance, in the case of $d=0$ (so $n=0$)
\begin{equation}
	\deg(V) = \frac{\ii}{2\pi} \int \tr(V^{-1}(\varphi) V'(\varphi)) \,\dd\varphi
\end{equation}
so the map $S^1 \to U(1)$ defined by $f(\varphi) = 1$ has no winding, whereas the map defined by $g(\varphi) = \ee^{-\ii \varphi}$ has a winding $\deg(g) = 1$. 
In our case, $M=\BZ \times S^1$ and we may use the first $2n$ variables to describe the Brillouin zone, $x_1 = k_1$ to $x_{2n} = k_{2n} = k_{d}$ and the last one for the time, $x_{2n+1} = t$. As an example, in $d=2$ (so $n=1$), the invariant is expressed as
\begin{equation}
	\begin{split}
	W_{\varepsilon}[U] &= \frac{1}{24 \pi^2} \int_{\BZ \times S^1} \, \dd k_x \dd k_y \dd t \\
	&\tr\left( 3 [V_{\varepsilon}^{-1} \partial_{k_x} V_{\varepsilon},V_{\varepsilon}^{-1} \partial_{k_y} V_{\varepsilon}] V_{\varepsilon}^{-1} \partial_t V_{\varepsilon} \right)
	\end{split}
\end{equation}
Note that if $D$ is a diffeomorphism on the manifold $M$, then $\deg(V \circ D) = \pm \deg(V)$ where the positive (negative) sign applies if $D$ is an orientation-preserving (orientation reversing) diffeomorphism.

Let $\varepsilon$ and $\varepsilon'$ be two quasi-energies and $P_{\varepsilon,\varepsilon'}(k)$ the spectral projector on states with quasi-energy between $\varepsilon$ and $\varepsilon'$ (to be precise, for $- 2\pi < \varepsilon_1, \varepsilon_2 < 0$, $P_{\varepsilon_1,\varepsilon_2}(k)$ is the spectral projector on eigenvalues  $\ee^{-\ii \varepsilon}$ in the arc joining $\ee^{-\ii  \varepsilon_1}$ and $\ee^{-\ii  \varepsilon_2}$ clockwise on the circle $U(1)$). The difference between $W$'s is related to a Chern invariant as
\begin{equation}
	\label{relation_W_chern}
	W_{\varepsilon'}[U] - W_{\varepsilon}[U] = C_{n}(\mathcal{E}_{\varepsilon, \varepsilon'}).
\end{equation}
Here $\mathcal{E}_{\varepsilon, \varepsilon'}$ is the vector bundle over $\BZ$ with fiber $P_{\varepsilon,\varepsilon'}(k) \mathbb{C}$ over $k\in \BZ$, corresponding to the quasi-energy band between $\varepsilon$ and $\varepsilon'$; and $C_{n}$ is the $n$-th Chern invariant\footnote{Following \cite{AvronSadunSegertSimon1989} we define the Chern invariants by integrating the Chern character forms (obtained from $\tr(\exp(\ii / 2 \pi \, \Omega))$, where $\Omega = P (\dd P)^2 P$ is the curvature) over the manifold, and not the Chern classes. When integrating the Chern classes, one would obtain the so-called Chern numbers $\tilde{C}_n \in \mathbb{Z}$ (see \cite{AzcarragaIzquierdo1998} for details). The Chern invariants are related to Chern numbers, in particular $C_1 = \tilde{C}_1$, and higher Chern invariants can be expressed in terms of the Chern numbers, but they do not necessarily coincide.}, computed from the projector family $P_{\varepsilon,\varepsilon'}$ as $C_{n}(P_{\varepsilon,\varepsilon'})$, where \cite{AvronSadunSegertSimon1989}
\begin{equation}
	\label{def_chern}
	C_{n}(P) = \int_{\BZ} \frac{1}{n!} \, \left( \frac{\ii}{2\pi} \right)^{n}  \tr\left[ P (\dd P)^{2n} \right].
\end{equation}

As the degree is additive, namely $\deg(u \cdot v) = \deg(u) + \deg(v)$ (provided all quantities are well-defined), we have
\begin{equation}
	W_{\varepsilon'}[U] - W_{\varepsilon}[U] = \deg(V_{\varepsilon'}) - \deg(V_{\varepsilon}) = \deg([V_{\varepsilon}]^{-1}V_{\varepsilon'}).
\end{equation}
Moreover, the difference between effective Hamiltonians at different gaps is equal to the spectral projector on the quasi-energy band between these gaps, $H^{\text{eff}}_{\varepsilon'}-H^{\text{eff}}_{\varepsilon} = (2\pi/T) P_{\varepsilon,\varepsilon'}$ so
\begin{equation}
	\label{VVP}
	[V_{\varepsilon}(t,k)]^{-1} V_{\varepsilon'}(t,k) = \ee^{2 \pi \ii t/T P_{\varepsilon,\varepsilon'}(k)}
\end{equation}
and the winding of this last unitary map can be shown to be the Chern invariant of projector family $P_{\varepsilon,\varepsilon'}$ (this is detailed in Appendix \ref{app_winding_chern}).

\section{Class AIII : with chiral symmetry}
\label{sec_classAIII}

In the following, we discuss what happens when chiral symmetry enters the game. On the one hand, when the dimension of space is even, the $W$ invariants at $\varepsilon$ and at $-\varepsilon$ are opposite when chiral symmetry is present; in the special gaps (that we dub \emph{chiral gaps}) around $0$ and $\pi$ (if they exist), $W$ vanishes. On the other hand, when the dimension of the space is odd, chiral symmetry enables to define a $\mathbb{Z}$ gap invariant characterizing only the chiral gaps, as it was done in the one-dimensional case \cite{AsbothTarasinskiDelplace2014}.

\subsection{Chiral symmetry and chiral basis}

\subsubsection{Chiral symmetry}

Chiral symmetry is usually presented as the product of particle-hole and time-reversal; there arise situations in which neither time-reversal nor particle-hole symmetries are preserved, yet chiral symmetry is. In static systems (as opposed to driven systems), chiral symmetry is present when the unitary chirality operator $\Gamma$ anticommutes with the Hamiltonian, that is, $\left\{H,\Gamma\right\} = 0$. The chirality operator $\Gamma \in U(N)$ is hermitian ($\Gamma^\dagger = \Gamma$) and can be chosen so that $\Gamma^2 = \Id$. Chiral symmetry can for example occur when the hoppings 
of the tight-binding model are bipartite \cite{Chiu2015}. Note that chiral symmetry is not a \emph{symmetry} in the usual sense as it anticommutes with the Hamiltonian instead of commuting with it \cite{Zirnbauer2004}; however, in the context of condensed matter (as well as systems such as mechanical, photonic or cold atoms lattices) it is meaningful to allow this possibility; we refer the reader to \cite[\S~3]{FreedMoore2013} for more details. Chiral symmetry has also been interpreted as underlying a supersymmetric structure \cite{Witten1981}, where two anticommuting supercharges $\mathcal{Q}_1 = H$ and $\mathcal{Q}_2 = \ii/2 [H,\Gamma] = \ii H \Gamma$ (combined into $\mathcal{Q}_{\pm} = (\mathcal{Q}_1 \pm \ii \mathcal{Q}_2)/2$) give rise to a supersymmetric Hamiltonian $\mathcal{H} = \mathcal{Q}_1^2 = \mathcal{Q}_2^2 = H^2$ (alternatively, $\left\{\mathcal{Q}_{+},\mathcal{Q}_{-}\right\} = \mathcal{H}$ and $\left\{\mathcal{Q}_{\pm},\mathcal{Q}_{\pm}\right\} = 0$) (see \cite{Ezawa2008,Kailasvuori2009} in electronic condensed matter and \cite{KaneLubensky2013,VitelliUpadhyayaGinGeChen2014,Lawler2015} in mechanical systems), an interpretation allowing the use of the whole toolbox of supersymmetry. The fate of this structure in the driven problem is an open question.

\subsubsection{The chiral basis}
\label{base_chirale}

As $\Gamma$ is unitary, it can always be diagonalized, and as $\Gamma^2 = \Id$ its eigenvalues are $\pm 1$, so there is a so-called \emph{chiral basis} where the matrix representing operator $\Gamma$ has the form
\begin{equation}
	\Gamma \cong \begin{pmatrix}
		\Id_{\gamma} & 0 \\
		0 & - \Id_{N - \gamma}
	\end{pmatrix} \in U(N).
\end{equation}
Despite the existence of several such bases, we will choose one and refer to it as \emph{the} chiral basis, as nothing actually depends on this choice. In the following, we will suppose, in order to simplify the discussion, that $N = 2 M$ is even, and that $\gamma = M$. However, the construction should be identical when this is not the case, up to the replacement of inverses of square matrices by conjugate transposes of rectangular matrices.

Now consider an operator $X$ such that
\begin{equation}
	\label{chiral_condition_mat}
	\left\{ \Gamma, X \right\} = 0
	\qquad
	\text{i.e.}
	\qquad
	\Gamma X \Gamma = - X
\end{equation}
and let us write the block matrix of this operator in the chiral basis
\begin{equation}
	X \cong \begin{pmatrix}
		A & B \\ C & D
	\end{pmatrix}.
\end{equation}
By computing the block product,
\begin{equation}
	\Gamma X \Gamma \cong \begin{pmatrix}
		A & -B \\ -C & D
	\end{pmatrix}
\end{equation}
so that \eqref{chiral_condition_mat} leads to the condition
\begin{equation}
	0 = \Gamma X \Gamma + X = \begin{pmatrix}
		2 A & 0 \\ 0 & 2 D
	\end{pmatrix}
\end{equation}
and therefore $A=0$ and $D=0$, so an operator that anticommutes with the chirality operator is block-antidiagonal in the chiral basis. In the same vein, an operator that commutes with the chirality operator is block-diagonal in the chiral basis.

\subsection{Chiral Hamiltonian and evolution operator}

Let us consider an arbitrary time-dependent Hamiltonian $H(t)$. This Hamiltonian is \emph{chiral} when
\begin{equation}
	\label{chiralite_hamiltonien}
	\Gamma H(t,k) \Gamma = - H(-t,k).
\end{equation}
More precisely, we only require that  $\Gamma H(t_0+t,k) \Gamma = - H(t_0-t,k)$ for some reference time $t_0$, but without loss of generality we may set $t_0 = 0$ (this is possible as long as there are no other symmetry or constraint). 

We now want to identify the constraint put by chirality on the evolution operator $U$, solution of the differential equation $\ii \dot{U} = H U$ with initial value $U(0) = \Id$. With this aim in mind, let us rephrase the previous condition on the Hamiltonian as $\Gamma H \Gamma = - H \circ \tau$ with $\tau(t,k) = (-t,k)$. Therefore, $\ii \Gamma \dot{U} \Gamma = \Gamma H \Gamma \, \Gamma U \Gamma$ as $\Gamma^2 = \Id$, that is to say, $\ii \dot{R} = (-H \circ \tau) R$ with $R = \Gamma U \Gamma$. Now define $S = U \circ \tau$ so $\ii \dot{S} = - \ii \dot{U} \circ \tau = - (H \circ \tau) (U \circ \tau) = (- H \circ \tau) S$. As the initial values $R(0) = S(0) = U(0) = \Id$ are the same, thanks to the uniqueness of the solution to the Cauchy problem we can identify $R=S$, that is to say $\Gamma U \Gamma = U \circ \tau$ or
\begin{equation}
	\label{chiralite_operateur_evolution}
	\Gamma U(t,k) \Gamma = U(-t,k).
\end{equation}

\subsection{Chirality and Floquet theory}

The aim of this paragraph is to determine the constraint put by the chiral symmetry \eqref{chiralite_hamiltonien} on the effective Hamiltonian, and on the periodized evolution operator when dealing with a periodically driven system.

First, remark that equality \eqref{chiralite_operateur_evolution} implies that
\begin{equation}
	\label{chiral_UFloquet}
	\Gamma U(T,k) \Gamma = U^{-1}(T,k)
\end{equation}
so the quasi-energy spectrum (laying on the unit circle of the complex plane) is let invariant by complex conjugation, i.e. by the reflection with respect to the real axis (as is visible on figures \ref{figure_ssh_dynamic_pi_nontrivial} and \ref{figure_ssh_dynamic_anomalous}). 

To determine the constraint put on the entire evolution by chiral symmetry, let us start with the effective Hamiltonian. With the definition \eqref{def_log} of the complex logarithm, we have
\begin{equation}
	\log_{-\varepsilon}(\ee^{-\ii \varphi}) = - \log_{\varepsilon}(\ee^{\ii \varphi}) - 2 \pi \ii.
\end{equation}
Therefore (omitting $k$), from \eqref{chiral_UFloquet} and \eqref{Heff_definition} we get
\begin{align}
 	\Gamma H^{\text{eff}}_{\varepsilon} \Gamma 
 	&= \frac{\ii}{T} \sum_{j} \log_{- \varepsilon}(\lambda_j) \Gamma\ket{\psi_j}\,\bra{\psi_j} \Gamma \\
	&= \frac{\ii}{T} \sum_{j} \log_{- \varepsilon}(\lambda_j^{-1}) \ket{\psi_j}\,\bra{\psi_j} \\
 	&= \frac{\ii}{T} \sum_{j} \left[ - 2 \pi \ii - \log_{\varepsilon}(\lambda_j) \right] \ket{\psi_j}\,\bra{\psi_j} \\
 	&= - H^{\text{eff}}_{-\varepsilon} + 2 \pi / T \; \Id
\end{align}
and at the end
\begin{equation}
	\label{chiralite_effective_hamiltonian}
	\Gamma H^{\text{eff}}_{\varepsilon} \Gamma 
	= - H^{\text{eff}}_{-\varepsilon} + \frac{2 \pi}{T} \Id.
\end{equation}
From this identity, one recovers the already mentioned reflection symmetry of the quasi-energy spectrum.

Now let us move to the periodized evolution operator defined in \eqref{periodized_evolution_operator_def}. One can use a series expansion of the exponential (along with the identity $\Gamma^2 = \Id$) to show that
\begin{equation}
	\Gamma \ee^{\ii t H^{\text{eff}}_{\varepsilon}} \Gamma = \ee^{\ii t \Gamma H^{\text{eff}}_{\varepsilon} \Gamma}
\end{equation}
which with~\eqref{chiralite_effective_hamiltonian} implies
\begin{equation}
	\Gamma \ee^{\ii t H^{\text{eff}}_{\varepsilon}} \Gamma = \ee^{-\ii t H^{\text{eff}}_{-\varepsilon}} \ee^{2 \pi \ii t/T}.
\end{equation}
As a consequence,
\begin{align}
	\Gamma V_{\varepsilon}(t, k) \Gamma &= \Gamma U(t,k) \Gamma \, \Gamma \ee^{\ii t H^{\text{eff}}_{\varepsilon}(k)} \Gamma \\
	&= U(-t,k) \ee^{-\ii t H^{\text{eff}}_{-\varepsilon}(k)} \ee^{2 \pi \ii t/T}
\end{align}
so we end up with the constraint
\begin{equation}
	\label{chiralite_operateur_evolution_periodise}
	\Gamma V_{\varepsilon}(t,k) \Gamma = V_{-\varepsilon}(-t,k) \ee^{2 \pi \ii t/T}.
\end{equation}

\subsection{Consequences of chiral symmetry on the $W$ invariant}

Constraint~\eqref{chiralite_operateur_evolution_periodise} implies that
\begin{equation}
	\deg(V_{\varepsilon}) = - \deg(V_{-\varepsilon}).
\end{equation}
This equality arises from the invariance of forms $\omega_{2n+1}$ under conjugation by the unitary operator $\Gamma$ and from the orientation-reversing character of the diffeomorphism $\tau(t,k) = (-t,k)$. In terms of the topological invariant $W$, we can state the following: when $U$ satisfies chiral symmetry \eqref{chiralite_operateur_evolution}, then
\begin{equation}
	W_{\varepsilon}[U] = - W_{-\varepsilon}[U].
\end{equation}
In particular, $W_0[U]$ and $W_{\pi}[U]$ necessarily vanish. However, other gaps (e.g. in a system with more than two bands) may still be topologically nontrivial: chiral symmetry gives a relation between them, but does not enforce their vanishing. This is, indeed, also the case in equilibrium systems where only the chiral gap at zero energy is trivial in presence of chiral symmetry\footnote{A somehow artificial but very simple example of this is the following. Consider a two-band Hamiltonian $H_0(k_x,k_y)$ realizing the anomalous quantum Hall effect, e.g. from the Haldane model \cite{Haldane1988}, and define $H = H_0 \otimes \sigma_z + \Delta \; \Id_2 \otimes \sigma_z$ with chiral symmetry given by $\Gamma = \Id_2 \otimes \sigma_x$. Set the parameter $\Delta$ so that is is larger that the bandwidth of $H_0$. The resulting Hamiltonian is chiral symmetric, $\{H,\Gamma\}=0$, it has four bands, the innermost gap being trivial whereas the two intermediate ones are topological and characterized with opposite $W$.}.

\subsection{Dynamical invariant for chiral systems}

\subsubsection{Preliminaries}

The constraint~\eqref{chiralite_operateur_evolution_periodise} is not particularly helpful to define a topological invariant specific to chiral systems because it relates operators at opposite times and corresponding to opposite gaps (or cuts $\varepsilon$). Let us first address the issue of opposite times: the periodicity of $V$ implies that $V_{\varepsilon}(-t,k) = V_{\varepsilon}(T-t,k)$ so at half-period (in $t=T/2$),
\begin{equation}
	\Gamma V_{\varepsilon}(T/2,k) \Gamma = - V_{-\varepsilon}(T/2,k).
\end{equation}
To shorten notations, we will omit the variable $k$ in the following.
In a periodically driven system, chiral symmetry relates states at opposite quasi-energies, so only the two values $\ee^{-i \varepsilon} = \pm 1$, corresponding to arguments $\varepsilon = \text{$0$ or $\pi$}$ are relevant. The case $\varepsilon = 0$ is obvious, but in the case of $\varepsilon = \pi$ we need the identity
\begin{equation}
	\log_{-(\varepsilon-2\pi)}(\ee^{\ii \varphi}) = \log_{-\varepsilon}(\ee^{\ii \varphi}) + 2 \pi \ii.
\end{equation}
so $V_{\varepsilon-2\pi}(T/2) =  - V_{\varepsilon}(T/2)$ and in particular $V_{-\pi}(T/2) =  - V_{\pi}(T/2)$.

To conclude, only for the gaps around $\varepsilon = 0$ or $\pi$ can chiral symmetry be used to define a topological invariant\footnote{In presence of chiral symmetry, we identified as symmetry of the quasi-energy spectrum the reflection by the real axis (going from $\ee^{\ii \varepsilon}$ to $\ee^{-\ii \varepsilon}$). The chiral gaps correspond to quasi-energies that are left invariant by this operation.}, and in this case
\begin{equation}
	\label{eq_action_gamma_vhalf}
	\begin{split}
	\Gamma V_{0}(T/2) \Gamma &= - V_{0}(T/2) \\
	\Gamma V_{\pi}(T/2) \Gamma &= + V_{\pi}(T/2).
	\end{split}
\end{equation}

Note that one could relate operators describing the same gap~$\varepsilon$, not necessarily $0$ or $\pi$, through chiral symmetry, but generically the phase $\ee^{2 \pi \ii t/T}$ would need to be replaced by a discontinuous operator accounting for the phase jumps of the logarithm, and that would therefore not be suitable to define a topological invariant.

\subsubsection{Case $\varepsilon = 0$}

Let us first deal with the case when $\varepsilon = 0$. In the chiral basis (see paragraph~\ref{base_chirale})
\begin{equation}
	\Gamma \cong \begin{pmatrix}
		\Id_{M} & 0 \\
		0 & - \Id_{M}
	\end{pmatrix}.
\end{equation}
In this basis, the periodized evolution operator at half-period is block-antidiagonal
\begin{equation}
	\label{antidiag_V0}
	V_{0}(T/2) \cong \begin{pmatrix}
		0 & V_{0}^{+} \\
		V_{0}^{-} & 0
	\end{pmatrix}
\end{equation}
where $V_{0}^{\pm} \in U(M)$ are unitary operators. The inverse matrix is
\begin{equation}
	[V_{0}(T/2)]^{-1} \cong \begin{pmatrix}
		0 & [V_{0}^{-}]^{-1} \\
		[V_{0}^{+}]^{-1} & 0
	\end{pmatrix}.
\end{equation}

\subsubsection{Case $\varepsilon = \pi$}

When $\varepsilon = \pi$, the periodized evolution operator at half-period is now block-diagonal,
\begin{equation}
	\label{diag_Vpi}
	V_{\pi}(T/2) \cong \begin{pmatrix}
		V_{\pi}^{+} & 0 \\
		0 & V_{\pi}^{-}
	\end{pmatrix}
\end{equation}
where again $V_{\pi}^{\pm} \in U(M)$, and
\begin{equation}
	[V_{\pi}(T/2)]^{-1} \cong \begin{pmatrix}
		[V_{\pi}^{+}]^{-1} & 0 \\
		0 & [V_{\pi}^{-}]^{-1}
	\end{pmatrix}.
\end{equation}

\subsubsection{Defining the invariant}

When the space and therefore the Brillouin torus are odd-dimensional (let $d=2n+1$ be this odd dimension), the winding of the periodized evolution operator at fixed time, $w(t)=\deg(k \mapsto V_{\varepsilon}(t, k))$, is well-defined, but vanishes. (Note that there are no unitary winding in even dimension, so it is not possible to proceed as in section \ref{sec:classA}.) Indeed, this degree is homotopy invariant and $V$ is smooth, so $w(t)$ does not actually depend on time; as $V_{\varepsilon}(t=0,k) = \Id$ for any $k$, $w(0) = 0$ and so $w(t) = 0$ for any time $t$. In particular, this is the case for $t=T/2$, which we shall use in the following.

We can use chiral symmetry to circumvent the vanishing of the invariant: as it will turn out, the two unitary submatrices that we identified in the previous section cancel each other in $w(t)$. However, we can use the block-(anti)diagonal structure of $V_{\varepsilon}(T/2)$ (equations \eqref{antidiag_V0} and \eqref{diag_Vpi} respectively for $\varepsilon = 0$ and $\varepsilon = \pi$) to define two unitary windings $\deg(V_{\varepsilon}^{\pm})$ from the periodized evolution operator. Computing $w(T/2)$ from the block-(anti)diagonal form of $V_{\varepsilon}(T/2)$, we see that
\begin{equation}
	0 = w(T/2) = \deg(V_{\varepsilon}^{+}) + \deg(V_{\varepsilon}^{-}),
\end{equation}
so there is actually only one independent invariant, which we denote by
\begin{equation}
	\label{chiral_invariant}
	G_{\varepsilon}[U] = \deg(V_{\varepsilon}^{+}).
\end{equation}
This is the general topological index for a chiral symmetric Floquet system in odd space dimension. Let us emphasize again that this invariant is only defined for gaps $\varepsilon = 0$ or $\pi$. 

The map $k \mapsto V_{\varepsilon}^{+}(k)$ does not depend on time (it is computed from the periodized evolution operator $V_{\varepsilon}$ at half period), so the degree in \eqref{chiral_invariant} is computed as an integral over the Brillouin torus only. For example, in $d=1$
\begin{equation}
	G_{\varepsilon}[U] = \frac{\ii}{2 \pi} \int_{\BZ} \tr\left( (V_{\varepsilon}^{+})^{-1} \partial_k V_{\varepsilon}^{+} \right) \, \dd k
\end{equation}
and in $d=3$,
\begin{equation}
	\begin{split}
	&G_{\varepsilon}[U] = \frac{1}{24 \pi^2} \int_{\BZ} \, \dd k_x \dd k_y \dd k_z \\
	&\tr\left( 3 [(V_{\varepsilon}^{+})^{-1} \partial_{k_x} V_{\varepsilon}^{+},(V_{\varepsilon}^{+})^{-1} \partial_{k_y} V_{\varepsilon}^{+}] (V_{\varepsilon}^{+})^{-1} \partial_{k_z} V_{\varepsilon}^{+} \right)
	\end{split}
\end{equation}
As $V_{\varepsilon}$ is computed both from the evolution operator $U$ and the effective Hamiltonian $H^{\text{eff}}_{\varepsilon}$, the invariant $G_{\varepsilon}$ still depends on the whole evolution (and not only of the state of the system at half period).

\subsection{Relation with the band invariant}

We now seek to relate the \emph{gap invariant} $G_{\varepsilon}[U]$ to the \emph{band invariant} used to characterize equilibrium systems computed for a quasi-energy band. In analogy with the case without symmetry, we expect this band invariant to be equal to the difference between the gap invariants in the two gaps surrounding the band; as we will show, this is indeed the case.

\subsubsection{Topological invariant for static chiral systems}

Let us first remind the reader of the construction of the chiral invariant for static systems (following the review \cite[p.~15]{Chiu2015}): a static chiral Hamiltonian such that
\begin{equation}
	\left\{ H, \Gamma \right\} = 0
\end{equation}
(with $\tr \Gamma = 0$ so the dimension of the state space is even, $N = 2M$) reads in the chiral basis
\begin{equation}
	H(k) \cong \begin{pmatrix}
		0 & D(k) \\ D^\dagger(k) & 0
	\end{pmatrix} \in \text{Herm}(2 M).
\end{equation}
The Hamiltonian is assumed to describe an insulator, so the valence (or conduction) projector $P(k)$ is well-defined, and is naturally associated to the unitary operator $Q(k) = \ee^{\ii \pi P(k)} = \Id - 2 P(k)$, which is also block-antidiagonal in the chiral basis,
\begin{equation}
	Q(k) \cong \begin{pmatrix}
		0 & q(k) \\ q^{\dagger}(k) & 0
	\end{pmatrix} \in U(2 M)
\end{equation}
where $q(k) \in U(M)$ is also unitary. The chiral invariant for the valence (or conduction) band is then defined as
\begin{equation}
	g[P] = \deg(q) \in \mathbb{Z}.
\end{equation}

\subsubsection{Relation between the gap invariants and the band invariants}

We now go back to our aim of showing that 
\begin{equation}
	\label{relation_gap_band_invariants_chiral}
	G_{\varepsilon}[U] - G_{\varepsilon'}[U] = g[P_{\varepsilon' \varepsilon}]
\end{equation}
where $P_{\varepsilon,\varepsilon'}(k)$ is the spectral projector on states with quasi-energy between $\varepsilon$ and $\varepsilon'$. As there are only two possible values for $\varepsilon$ or $\varepsilon'$, it is sufficient to show that
\begin{equation}
	G_{0}[U] - G_{\pi}[U] = g[P_{0 \pi}].
\end{equation}
As we already used in the first part (in equation \eqref{VVP}),
\begin{equation}
	[V_{\pi}(t,k)]^{-1} V_{0}(t,k) = \ee^{2 \pi \ii t/T P_{\pi,0}(k)}
\end{equation}
so
\begin{equation}
	\label{relation_vpi_v0_Qpi0}
	[V_{\pi}(T/2,k)]^{-1} V_{0}(T/2,k) = \ee^{\ii \pi P_{\pi,0}(k)} = Q_{\pi,0}(k).
\end{equation}

From equations \eqref{relation_vpi_v0_Qpi0} and \eqref{eq_action_gamma_vhalf} we infer that $Q_{\pi,0}$ is block-antidiagonal in the chiral basis, namely
\begin{equation}
	Q_{\pi,0}(k) \cong \begin{pmatrix}
		0 & q_{\pi,0}^{+}(k) \\ q_{\pi,0}^{-}(k) & 0
	\end{pmatrix} \in U(2 M).
\end{equation}
Besides, we can compute the block product 
\begin{equation}
\begin{split}
	&[V_{\pi}(T/2,k)]^{-1} V_{0}(T/2,k) \\
		&=\begin{pmatrix}
		0 & [V^{+}_{\pi}(k)]^{-1} V^{+}_{0}(k) \\ [V^{-}_{\pi}(k)]^{-1} V^{-}_{0}(k) & 0
	\end{pmatrix}
\end{split}
\end{equation}
and therefore,
\begin{equation}
	[V^{+}_{\pi}(k)]^{-1} V^{+}_{0}(k) = q_{\pi,0}^{+}(k).
\end{equation}
As the degree \eqref{def_winding} is additive, we end up with
\begin{equation}
	\deg(V^{+}_{0}) - \deg(V^{+}_{\pi}) = \deg(q_{\pi,0}^{+})
\end{equation}
which is the identity \eqref{relation_gap_band_invariants_chiral} that we wanted to show.

\section{An example with chiral symmetry in 3D}

\begin{figure}[!thb]
	\centering
	\includegraphics{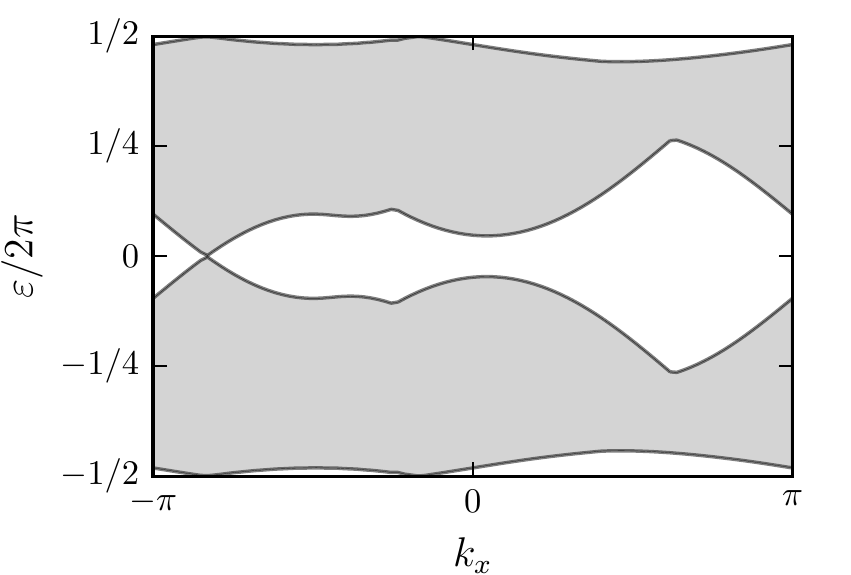}
	\vspace*{-0.5cm}
	\caption{\label{figure_3d_chiral_spectrum} Projection along $k_x$ of the quasi-energy spectrum of a finite system (with edges) of the 3D driven chiral model for $\delta = \num{1/2}$, $m_0 = \num{1.75}$, $m_1 = \num{1}$, and $\omega = \num{5}$. Bands are in gray. The driven system is topological, with $G_{0}[U] = \num{1}$ and $G_{\pi}[U] = \num{-2}$. As a consequence, surface states with linear dispersion (Dirac cones) appear in the quasi-energy spectrum of the finite system. In accordance with the bulk topological invariants, for each (bottom and top) surface boundary of the system, there are one Dirac cone in bulk gap $0$ and two Dirac cones in bulk gap $\pi$. The quasi-energy spectrum is obtained by diagonalization in the Sambe space truncated to \num{7} sidebands for a system of length \num{15}. The invariants were computed from the bulk Hamiltonian by direct integration (e.g. we obtain here $G_{0}=\num{0.99}$ and $G_{\pi}=\num{-1.97}$ which are rounded to the values given in the main text). The code used for all computations is available in Ref.~\bibnotemark[sourcecode].
	}
\end{figure}

\begin{figure}[!thb]
	\centering
	\includegraphics{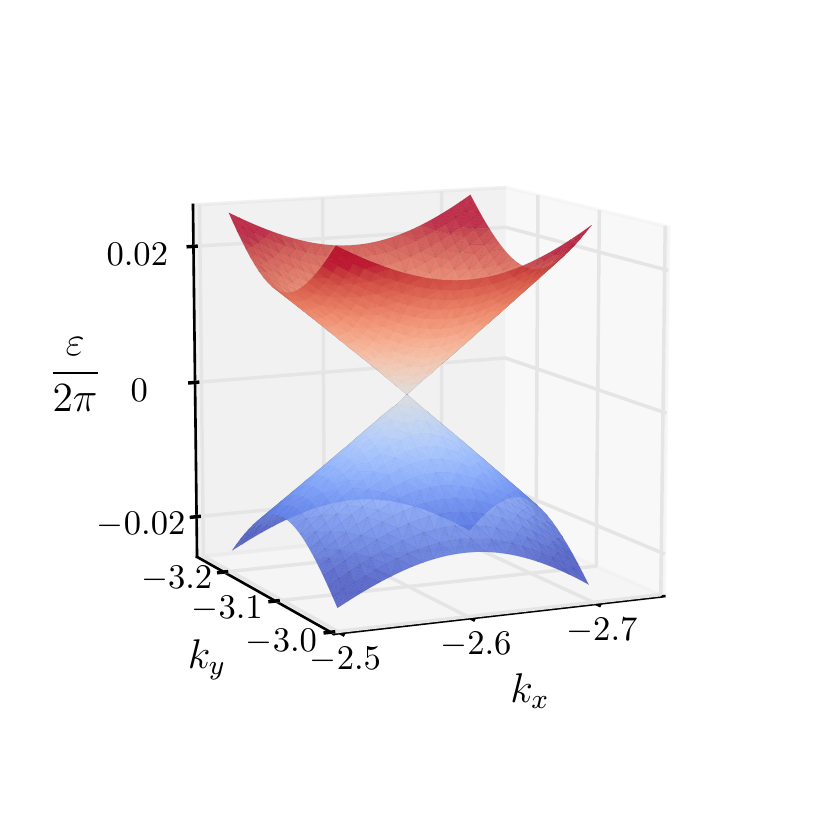}
	\vspace*{-0.5cm}
	\caption{\label{figure_3d_chiral_DP0_zoom} Three dimensional view of the Dirac cone in bulk gap~$0$ of the quasi-energy spectrum of figure~\ref{figure_3d_chiral_spectrum}. The dispersion is clearly linear in the neighborhood of the Dirac cone (located approximately at $k_x = \num{-2.62}$ and $k_y = - \pi$). The cone then merges in the bulk bands (shown in figure~\ref{figure_3d_chiral_spectrum}).
	}
\end{figure}

In section \ref{example_chiral}, we discussed the well-known driven SSH model. We will now illustrate the new invariants introduced in this article in a three dimensional driven tight-binding system with chiral symmetry, obtained by a periodic modulation of the parameters of a minimal tight-binding model for chiral topological insulators adapted from \cite{WangDengMooreSunDuan2015}. Consider electrons with spin $1/2$ on a crystal with two sublattices (or orbitals) $A$ and $B$, with Bloch Hamiltonian
\begin{equation}
	H(t,k) = \begin{pmatrix}
		0 & \ii h_0 \Id + h_i(t,k) \sigma_{i} \\
		[\ii h_0 \Id + h_i(t,k) \sigma_{i}]^{\dagger} & 0
	\end{pmatrix}
\end{equation}
in basis $(A \uparrow, A \downarrow, B \uparrow, B \downarrow)$, and where $h_0 = m(t) + \cos k_x + \cos k_y + \cos k_z$, $h_1 = \delta + \sin k_x$, $h_2 = \sin k_y$ and $h_3 = \sin k_z$. Chiral symmetry is represented by the matrix $\Gamma = \sigma_3 \otimes \sigma_0$ (where $\sigma_3$ acts on sublattice degrees of freedom whereas $\sigma_0$ acts on spin degrees of freedom).  
This Hamiltonian is chiral, $\Gamma H(t,k) \Gamma = - H(-t,k)$, provided that $m(t) = m(-t)$, so we take $m(t) = m_0 + m_1 \cos(\omega t)$. In the following, we fix $\delta = 1/2$. The undriven system with $m_1 = 0$ is a chiral topological insulator in class AIII, and has a nontrivial chiral (band) invariant $g \neq 0$ for $|m_0| < 1$ (where $g = - 2$)  and for $1 < |m_0| < 2$ (where $g = 1$) \cite{WangDengMooreSunDuan2015}. We set $m_0 = \num{1.75}$ so $g = \num{1}$ in the undriven system.  We expect the periodic driving to induce topological phase transitions; indeed, for a driving of angular frequency $\omega = 5$ and amplitude $m_1 = 1$, the numerical computation of the invariants defined in section \ref{sec_classAIII} gives $G_{0}=\num{1}$ and $G_{\pi}=\num{-2}$, a situation only possible in driven systems, as there are surface states both in the gaps at quasi-energy $\varepsilon = 0$ and at quasi-energy $\varepsilon = \pi$. 

The numerical computation of the quasi-energy spectrum of a finite system reveals the appearance of Dirac cones (that is, of isolated points on the Brillouin zone where the gap closes with linear dispersion) at the surface of the system. As expected from the bulk-boundary correspondence principle, we found that at each (bottom and top) interface, the bulk gap $\varepsilon$ hosts $G_{\varepsilon}$ Dirac cones. This is illustrated in figures~\ref{figure_3d_chiral_spectrum} and~\ref{figure_3d_chiral_DP0_zoom}.

\section*{Conclusion}

In this article, we extended to any dimension and put in a single framework the topological invariants for periodically driven bulk Hamiltonians of periodic crystals in the complex Altland-Zirnbauer classes A (no symmetry) and AIII (chiral symmetry). As expected, these invariants are of the same nature ($\mathbb{Z}$-valued) as the ones of static systems, but to fully describe the properties of driven systems, it is necessary to define invariants characterizing gaps and not bands. It is however possible to use gap invariants to characterize static systems, which can be more practical as the gap invariants directly count the number of chiral edge states in a finite system. Band invariants and gap invariants are related in the following way: the band invariant is the difference of the surrounding gap invariants. This property is conjectured to be general. 
We did not address the subject of weak invariants (that is, lower-dimensional invariants); for example, in a $3$-dimensional system, the one-dimensional chiral invariant is still well-defined in the bulk. We expect their consequences on finite systems to be the same as in static systems. 

\begin{acknowledgments}
	I am grateful to D. Carpentier, P. Deplace, K. Gawędzki and C. Tauber for continuous exchanges and valuable discussions.
\end{acknowledgments}

\appendix

\section{The Bott map between unitary windings and Chern invariants}
\label{app_winding_chern}

There is a profound relation between the windings of unitary maps and the $n^{\text{th}}$ Chern invariants of projector families. Let us consider the unitary map
\begin{equation}
	\label{bott_map_V_P}
	V(t,k) = \ee^{\ii t P(k)}
\end{equation}
where $P(k)$ is a projector over the $2n$-dimensional Brillouin zone $\BZ$ (and $k \in \BZ$) and $t \in [0,2\pi]$. We have then
\begin{equation}
	\label{relation_winding_chern}
	\deg_{2n+1}(V) = - C_{2n}(P).
\end{equation}
This identity can be understood from an abstract point of view as realizing the K-theoretical isomorphism \cite{RordamLarsenLaustsen2000}
\begin{equation}
	K^{0}(X) \simeq K^{1}(SX)
\end{equation}
where $X$ is a manifold and $SX$ is its suspension (which is essentially the cylinder $[0,2\pi] \times X$ where all end-points $(0,x)$ on the one hand and $(2\pi,x)$ on the other hand are separately identified). Here, $X=\BZ$ is the Brillouin torus, and the fact that both $V(0)$ and $V(T)$ are constant over the Brillouin torus ($V(0,k)=\Id=V(T,k)$ for all $k \in \BZ$), let us see them as functions on $SX$, this isomorphism being realized by the Bott map \eqref{bott_map_V_P}. In a nutshell, $K^{0}(X)$ provides a notion of equivalence between families of projector families on $X$ (or equivalently vector bundles over $X$) that is weaker than homotopy equivalence, but that persists when the projectors are included in a bigger space (in the context of band theory for example, we expect that a trivial band added at some high energy in the spectrum will not modify the topological properties of the system, as explained by Kitaev \cite{Kitaev2009}); $K^{1}(X)$ provides the same kind of weak homotopy equivalence, but for unitary families on $X$ (we refer the reader to e.g. \cite{RordamLarsenLaustsen2000} for details).

It is also possible to derive equality \eqref{relation_winding_chern} directly, as did \cite{RudnerLindnerBergLevin2013} is the case $d=2$. We want to show that when \enquote{integrating over time} the form $\omega_{2n+1}$, one recovers the form that Chern character which, integrated over the Brillouin torus, gives the Chern invariant. Let us consider the differential form
\begin{equation}
\chi = \tr\left[(V^{-1}\dd V)^{2n+1}\right]
\end{equation} 
for the specific case where $V$ is given by~\eqref{bott_map_V_P}. Singling out the time variable, 
\begin{equation}
	\chi = (2n+1) \tr\left[V^{-1}\partial_{t}V \dd t \wedge (V^{-1} \dd V)^{2n}\right]
\end{equation}
and only the spatial part of $(V^{-1} \dd V)^{2n}$, that is $(V^{-1} \dd_{\BZ} V)^{2n}$, contributes to $\chi$. Now compute (following \cite{RudnerLindnerBergLevin2013})
\begin{equation}
	V^{-1}\partial_{t}V = \ii P
\end{equation}
and 
\begin{equation}
	V^{-1} \dd_{\BZ} V = a(t) P \dd P + b(t) \dd P
\end{equation}
where $a(t) = 2 (1-\cos t)$ and $b(t) = \ee^{\ii t} - 1$. As $P^2 = P$ is a projector, $P (\dd P) P = 0$ so
\begin{equation}
	\tr\left[P (V^{-1} \dd_{\BZ} V)^{2n}\right] = b^{n} (a + b)^{n} \tr\left[ P (\dd P)^{2n} \right].
\end{equation}
Therefore, one has
\begin{equation}
	\iota_{\partial_t} \chi = (2n+1) \ii \; b^{n} (a + b)^{n} \; \tr\left[ P (\dd P)^{2n} \right]
\end{equation}
where $\iota_{\partial_t}$ is the interior product (which basically removes $\dd t$ from the differential form in a coherent way; for example if the form $\alpha$ does not contain $\dd t$, then $\iota_{\partial_t} \dd t \wedge \alpha = \alpha$).
Integrating over one time period,
\begin{equation}
	\int_{0}^{2 \pi}\!\!\!\!\dd t\; \iota_{\partial_t} \chi = (2n+1) \ii \; \mathcal{I}_n \; \tr\left[ P (\dd P)^{2n} \right]
\end{equation}
where
\begin{equation}
	 \mathcal{I}_n = \int_{0}^{2 \pi}\!\!\!\!  b^{n}(t) (a(t) + b(t))^{n} \dd t  = 2 \pi (-1)^n  \frac{(2n)!}{(n!)^2}
\end{equation}
so
\begin{equation}
	\int_{0}^{2 \pi}\!\!\!\! \dd t\; \iota_{\partial_t} \chi = \frac{2 \pi}{\ii} (-1)^{n+1}  \frac{(2n+1)!}{(n!)^2} \tr\left[ P (\dd P)^{2n} \right].
\end{equation}
Along with the definitions \eqref{def_chern} and \eqref{def_winding}, this proves that for the specific case of the map \eqref{bott_map_V_P}, we have
\begin{equation}
	\deg_{2n+1}(V) = - C_{2n}(P).
\end{equation}
Note that with the usual definitions of these topological quantities, there is an additional minus sign in this identity, which is only matter of convention and could be absorbed in a definition.

\bibliography{bibliographie_chiral_floquet}

\end{document}